\documentclass[epjST]{svjour}

\usepackage{graphics}

\newcommand{\ba}{\begin{eqnarray}}
\newcommand{\ea}{\end{eqnarray}}

\begin{document}

\title{Discrete Symmetries in the Cluster Shell Model}

\author{A.H. Santana Vald\'es \and R. Bijker\thanks{\email{bijker@nucleares.unam.mx}}}

\institute{Instituto de Ciencias Nucleares, 
Universidad Nacional Aut\'onoma de M\'exico,
A.P. 70-543, 04510 M\'exico D.F., M\'exico}

\abstract{The role of discrete (or point-group) symmetries is discussed in the framework 
of the Cluster Shell Model which describes the splitting of single-particle levels in the 
deformed field of cluster potentials. We discuss the classification of the eigenstates 
for the cases of a triangular and tetrahedral configuration of $\alpha$-particles in terms 
of the irreducible representations of the double point groups ${\cal D}'_{3h}$ and ${\cal T}'_d$, 
respectively, and show how the discrete symmetry of a given eigenstate can be determined. 
Finally, we derive the Coriolis coupling for each one of these geometrical configurations.}

\maketitle

\section{Introduction}

Discrete symmetries have been used in nuclear physics in the context of collective models 
to characterize the intrinsic shape of the nucleus, such as axial symmetry for quadrupole 
deformations \cite{BMVol2}, and tetrahedral \cite{Dudek1,Dudek2} and octahedral \cite{Dudek2,Piet} 
symmetries for deformations of higher multipoles. A different application of the concept of 
discrete symmetries is found in the context of $\alpha$-particle clustering in light nuclei 
to describe the geometric configuration of the $\alpha$-particles. Early work on $\alpha$-cluster 
models goes back to the 1930's with studies by Wheeler \cite{wheeler}, and Hafstad and Teller 
\cite{Teller}, followed by later work by Brink \cite{Brink1,Brink2} and Robson \cite{Robson}. 
The measurements in recent years of new rotational excitations of the ground state of $^{12}$C 
with $L^P=4^-$ and $5^-$ \cite{Fre07,Kirsebom,Daniel} and of the Hoyle state with $L^P=2^+$ and $4^+$ 
\cite{Itoh,Freer12,Gai,Fre11} have generated a large renewed interest in the structure of $^{12}$C,
and of $\alpha$-cluster nuclei in general \cite{FreerFynbo,Schuck,Freer}. 

The experimental verification of the existence of discrete symmetries in nuclei consists in the 
study of the structure of rotational bands as fingerprints of the underlying discrete symmetry 
involving both the angular momentum and parity content of rotational bands and electromagnetic 
transitions and moments \cite{C12-1,C12-2,O16-1,O16-2,Dudek}. 
An analysis of the available experimental data of $k \alpha$ nuclei has provided evidence 
for the existence of triangular ${\cal D}_{3h}$ symmetry in $^{12}$C and tetrahedral ${\cal T}_d$ 
symmetry in $^{16}$O \cite{Daniel,C12-1,C12-2,O16-1,O16-2,Bijker2016}. An interesting question is to what 
extent these geometric configurations are manifested in the neighboring odd-mass nuclei, and what 
are their characteristic signatures. Hereto the Cluster Shell Model 
(CSM\footnote{Not to be confused with the Cranked Shell Model}) has been developed 
which describes the splitting of single-particle levels in deformed cluster potentials 
\cite{CSM1} with applications to $^{9}$Be and $^{9}$B \cite{CSM2}, and $^{13}$C \cite{BI2019}. 

The aim of this contribution is to discuss the classification of the eigenstates in the CSM 
for the cases of a triangular and tetrahedral configuration of $\alpha$-particles, related 
to the double point groups ${\cal D}'_{3h}$ and ${\cal T}'_d$, respectively, and show how 
the discrete symmetry of a given eigenstate can be determined. As an application,  
we derive the Coriolis coupling for each one of these geometrical configurations.

\section{Cluster Shell Model}

The cluster shell model has been introduced recently \cite{CSM1,CSM2,PPNP} to describe nuclei composed 
of $k$ $\alpha$-particles plus additional nucleons, simply denoted as $k \alpha + x$ nuclei 
The CSM combines cluster and single-particle degrees of freedom, and is very similar in spirit as the 
Nilsson model \cite{Nilsson}, but in the CSM the odd nucleon moves in the deformed field generated by the 
(collective) cluster degrees of freedom. The Hamiltonian is written as
\ba
H \;=\; T + V(\vec{r}) + V_{\rm so}(\vec{r}) + \frac{1}{2}(1+\tau_3) V_{\rm C}(\vec{r}) ~,
\label{hcsm}
\ea
{\it i.e.} the sum of the kinetic energy, a central potential obtained by convoluting 
the density 
\ba
\rho(\vec{r}) &=& \left( \frac{\alpha}{\pi}\right)^{3/2} 
\sum_{i=1}^{k}\exp \left[ -\alpha \left( \vec{r}-\vec{r}_{i}\right)^{2}\right] 
\nonumber \\
&=& \left( \frac{\alpha }{\pi }\right)^{3/2} \mbox{e}^{-\alpha(r^{2}+\beta ^{2})} 
\, 4\pi \,\sum_{\lambda\nu} i_{\lambda}(2\alpha \beta r) Y_{\lambda\nu}(\theta,\phi) 
\sum_{i=1}^{k} Y_{\lambda\nu}^{\ast}(\theta_{i},\phi_{i}) ~,  
\label{rhor}
\ea
with the interaction between the $\alpha$-particle and the nucleon, a spin-orbit interaction and, 
for an odd proton, a Coulomb potential. Here $\vec{r}_i=(r_i,\theta_i,\phi_i)$ denotes the coordinates 
of the $\alpha$-particles with respect to the center-of-mass. In this manuscript we consider equilateral 
triangular and tetrahedral configurations for which the distance of the $\alpha$-particles with 
respect to the center-of-mass is the same $r_i=\beta$. 

\begin{figure}
\centering
\begin{minipage}{.5\linewidth}
\resizebox{1.1\columnwidth}{!}{\includegraphics{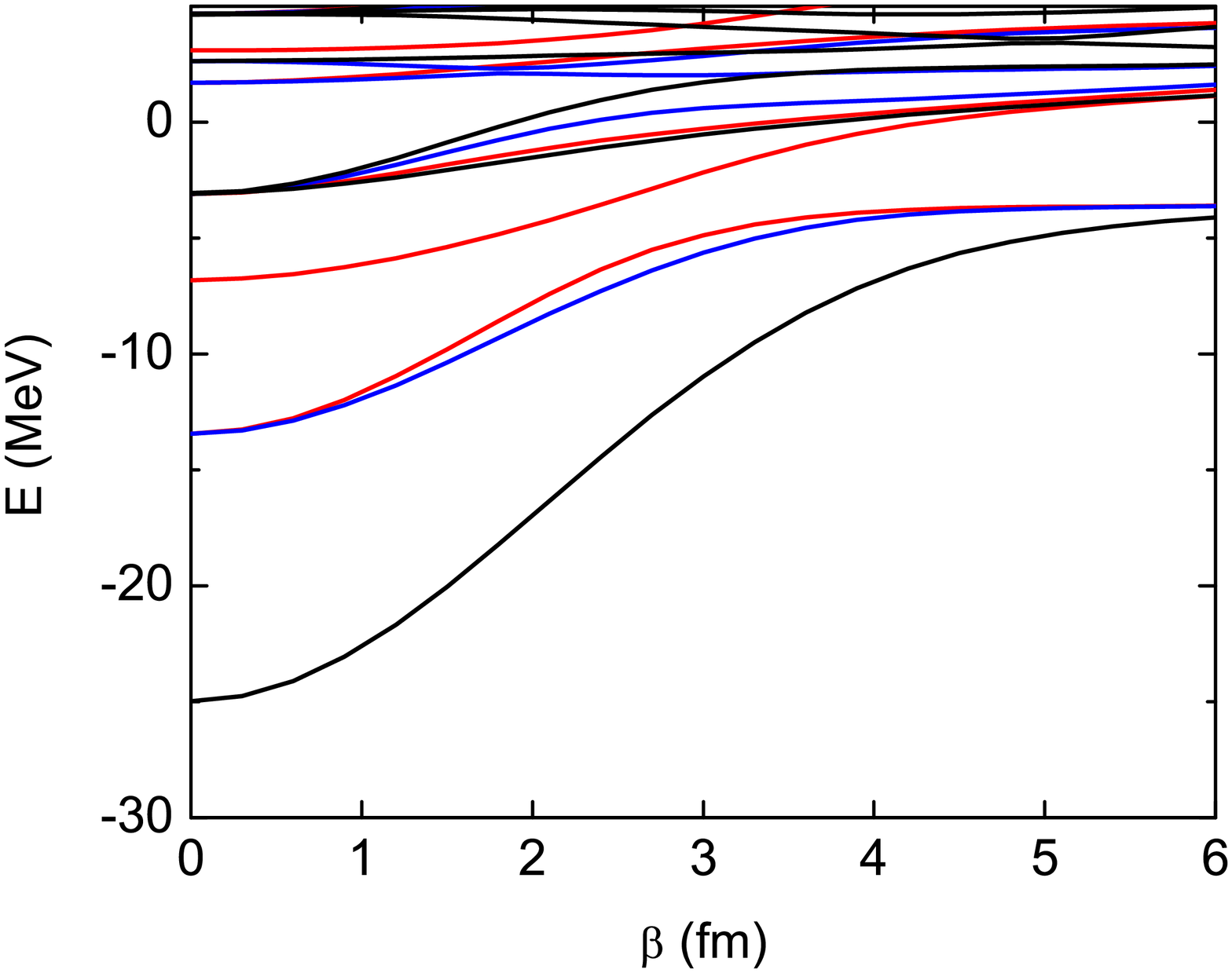}}
\end{minipage}\hfill
\begin{minipage}{.5\linewidth}
\resizebox{1.1\columnwidth}{!}{\includegraphics{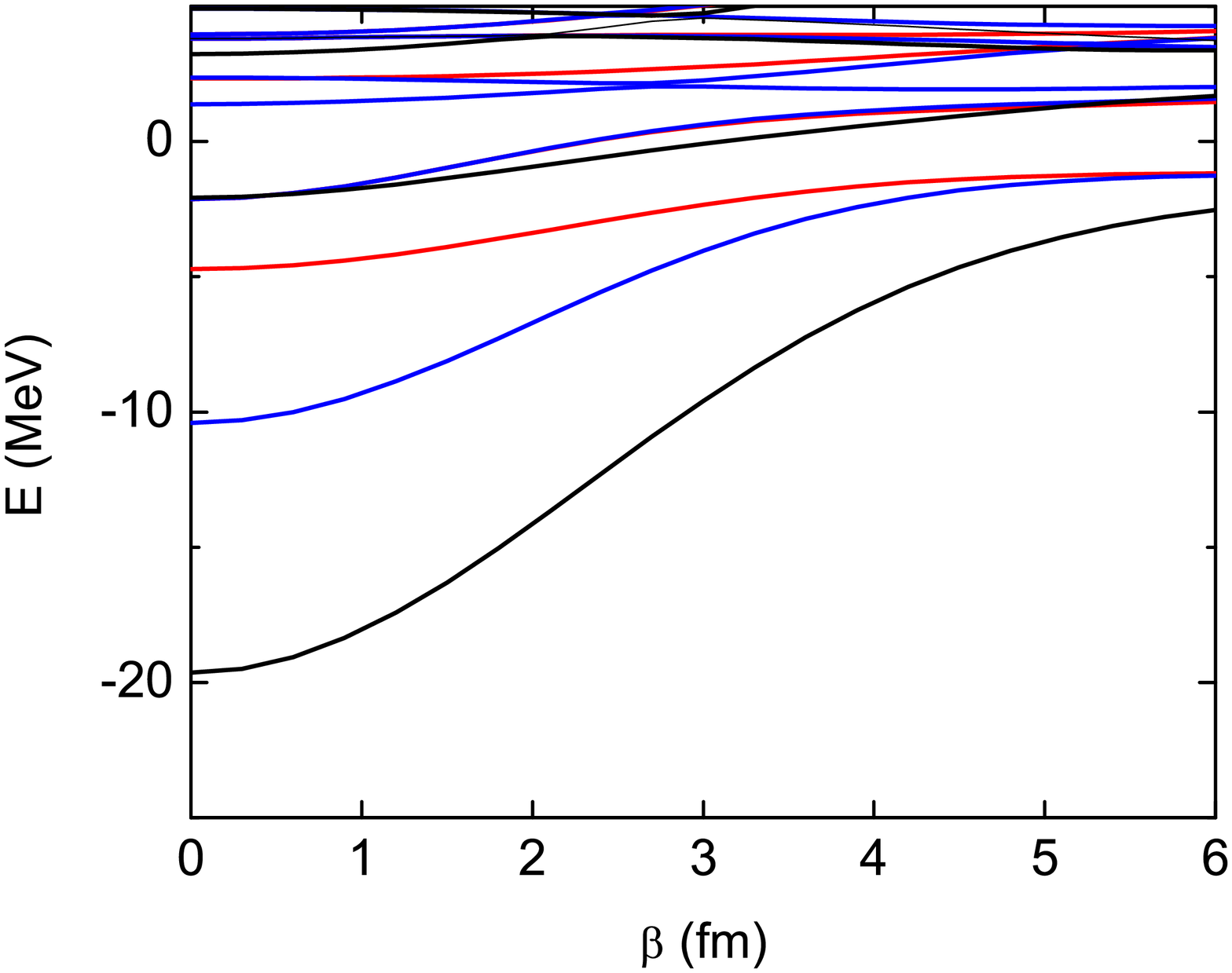}}
\end{minipage}\hfill
\caption{Single-particle energies in a cluster potential 
with ${\cal D'}_{3h}$ triangular symmetry (left) and 
with ${\cal T'}_d$ tetrahedral symmetry (right). 
In the left panel, the single-particle levels are labeled 
by $E_{1/2}$ (black), $E_{5/2}$ (red) and $E_{3/2}$ (blue), 
and in the right panel by $E_{1/2}$ (black), $E_{5/2}$ (red) 
and $G_{3/2}$ (blue). For $\beta=0$ the ordering of the single-particle 
orbits is $1s_{1/2}$, $1p_{3/2}$, $1p_{1/2}$ and (almost degenerate) $1d_{5/2}$, $2s_{1/2}$.}
\label{splevels}
\end{figure}

The Hamiltonian of the CSM of Eq.~(\ref{hcsm}) is solved in the intrinsic, or body-fixed, system. 
The single-particle energies and the intrinsic wave functions are obtained by diagonalizing 
the $H$ in the harmonic oscillator basis $| nljm \rangle$. In Fig.~\ref{splevels} we show the results 
for a single-particle moving in the deformed field generated by a cluster of $\alpha$-particles 
with triangular and tetrahedral symmetry as a function of $\beta$, the distance of the $\alpha$-particles 
with respect to the center-of-mass.

\section{Triangular symmetry: $D_{3h}$}

For a cluster of three identical $\alpha$-particles it 
is convenient to choose the $z$-axis along the symmetry axis of the triangle. 
In this geometry, the coordinates of the three $\alpha$-particles are given by 
$\vec{r}_1=(\beta,\pi/2,0)$, $\vec{r}_2=(\beta,\pi/2,2\pi/3)$ and 
$\vec{r}_3=(\beta,\pi/2,4\pi/3)$. Note, that this configuration is different 
from the one used in Ref.~\cite{CSM1}. With this choice, the geometric factor 
in the Hamiltonian becomes
\ba
\sum_{i=1}^{3} Y^{\ast}_{\lambda \nu}(\theta_i,\phi_i) 
\;=\; Y_{\lambda \nu}(\frac{\pi}{2},0) \left[ 1 + 2\cos (\frac{2\pi\nu}{3}) \right]
\;=\; 3\delta_{\nu,3\kappa} \, Y_{\lambda \nu}(\frac{\pi}{2},0) ~,
\label{triangle}
\ea
{\it i.e.} $\nu$ is a multiple of three. The eigenstates are intrinsic states upon 
which an entire rotational band is built. The structure of the rotational bands can 
be obtained by studying the point group symmetry of the problem. 

For the case of triangular symmetry the eigenstates of the CSM can be classified according 
to the irreducible representations (irreps) of the double point group ${\cal D}'_{3h}$, 
The double group ${\cal D}'_{3h}$ has three doubly degenerate spinor representations, 
denoted by Koster in applications to crystal physics as $\Gamma_{7}$, $\Gamma_{8}$, 
$\Gamma_{9}$ \cite{Koster} and by Herzberg in applications to molecular physics as $E_{1/2}$, 
$E_{5/2}$, $E_{3/2}$ \cite{Herzberg3}. In a recent application to nuclear physics the three 
representations were written as \cite{BI2019} $E_{1/2}^{(+)} \equiv \Gamma_{7} \equiv E_{1/2}$, 
$E_{1/2}^{(-)} \equiv \Gamma_{8}\equiv E_{5/2}$ and $E_{3/2} \equiv \Gamma_{9} \equiv E_{3/2}$.  
In this article, we adopt the notation of Herzberg.

\begin{table}[b]
\centering
\caption{Resolution of rotational states with half-integer angular momentum $J$ and parity $P$  
into irreps of ${\cal D}'_{3h}$ \cite{Koster,Herzberg3}.}
\label{d3hj}
\begin{tabular}{c|ccc|c|cccc}
\hline
\noalign{\smallskip}
& $E_{1/2}^{(+)}$ & $E_{1/2}^{(-)}$ & $E_{3/2}$ & 
& $E_{1/2}^{(+)}$ & $E_{1/2}^{(-)}$ & $E_{3/2}$ & \cite{BI2019} \\ 
\noalign{\smallskip}
$D_J^P$ & $\Gamma_7$ & $\Gamma_8$ & $\Gamma_9$ &
$D_J^P$ & $\Gamma_7$ & $\Gamma_8$ & $\Gamma_9$ & \cite{Koster} \\
\noalign{\smallskip}
& $E_{1/2}$ & $E_{5/2}$ & $E_{3/2}$ & 
& $E_{1/2}$ & $E_{5/2}$ & $E_{3/2}$ & \cite{Herzberg3} \\ 
\noalign{\smallskip}
\hline
\noalign{\smallskip}
$1/2^+$ & 1 & 0 & 0 & $1/2^-$ & 0 & 1 & 0 & \\
\noalign{\smallskip}
$3/2^+$ & 1 & 0 & 1 & $3/2^-$ & 0 & 1 & 1 & \\
\noalign{\smallskip}
$5/2^+$ & 1 & 1 & 1 & $5/2^-$ & 1 & 1 & 1 & \\
\noalign{\smallskip}
$7/2^+$ & 1 & 2 & 1 & $7/2^-$ & 2 & 1 & 1 & \\
\noalign{\smallskip}
$9/2^+$ & 1 & 2 & 2 & $9/2^-$ & 2 & 1 & 2 & \\
\noalign{\smallskip}
$11/2^+$ & 2 & 2 & 2 & $11/2^-$ & 2 & 2 & 2 & \\
\noalign{\smallskip}
$13/2^+$ & 3 & 2 & 2 & $13/2^-$ & 2 & 3 & 2 & \\
\noalign{\smallskip}
\hline
\end{tabular}
\end{table}

Table~\ref{d3hj} shows the resolution of angular momentum states into irreps of the double point 
group ${\cal D}'_{3h}$. The results for higher angular momenta ($J \ge 13/2$) are given by 
\ba
D_J^{P} \;=\; D_{J-6}^{P} + (2E_{1/2}+2E_{5/2}+2E_{3/2}) ~.
\ea
For the triangular configuration, the rotational states can be labeled by the angular momentum $J$ 
and its projection $K$ on the symmetry axis, $| J^P K \rangle$. Both $J$ and $K$ are half integer. 
The allowed values of $K^P$ for each one of the spinor representations are given by \cite{BI2019}
\ba
\begin{array}{lcl}
\Omega = E_{1/2} &: \qquad& K^P = \frac{1}{2}^+, \frac{5}{2}^-, \frac{7}{2}^-,  
\frac{11}{2}^+, \frac{13}{2}^+, \ldots \\ && \\
\Omega = E_{5/2} &: \qquad& K^P = \frac{1}{2}^-, \frac{5}{2}^+, \frac{7}{2}^+,  
\frac{11}{2}^-, \frac{13}{2}^-, \ldots \\ && \\
\Omega = E_{3/2} &: \qquad& K^P = \frac{3}{2}^{\pm}, \frac{9}{2}^{\pm}, \frac{15}{2}^{\pm}, \ldots 
\end{array}
\ea
The $K^P$ bands for the representation $E_{3/2}$ are doubly degenerate (parity doubling).
The angular momenta of each $K$ band are given by $J=K,K+1,K+2,\ldots$, to obtain
\ba
\begin{array}{lcl}
\Omega = E_{1/2} &: \qquad& J^P = \frac{1}{2}^+, \frac{3}{2}^+, \frac{5}{2}^{\pm}, 
\frac{7}{2}^+, (\frac{7}{2}^-)^2, \frac{9}{2}^+, (\frac{9}{2}^-)^2, \ldots \\ && \\
\Omega = E_{5/2} &: \qquad& J^P = \frac{1}{2}^-, \frac{3}{2}^-, \frac{5}{2}^{\pm}, 
\frac{7}{2}^-, (\frac{7}{2}^+)^2, \frac{9}{2}^-, (\frac{9}{2}^+)^2, \ldots \\ && \\
\Omega = E_{3/2} &: \qquad& J^P = \frac{3}{2}^{\pm}, \frac{5}{2}^{\pm}, \frac{7}{2}^{\pm}, 
(\frac{9}{2}^{\pm})^2, \ldots 
\end{array}
\label{angmom3}
\ea
in agreement with Table~\ref{d3hj}.

\begin{figure}
\centering
\resizebox{0.9\columnwidth}{!}{\includegraphics{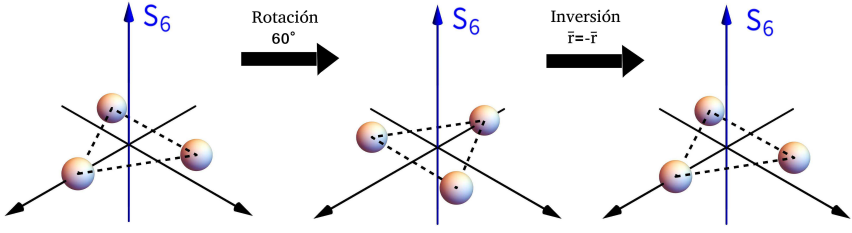}} 
\caption{Triplex symmetry.}
\label{Triplex}
\end{figure}

The eigenstates of the CSM Hamiltonian of Eq.~(\ref{hcsm}) can be classified according to the irreps 
of the ${\cal D}'_{3h}$ triangular symmetry as $| \Omega,\mu \rangle$, where $\Omega$ denotes each 
one of the doubly degenerate spinor representations $\Omega=E_{1/2}$, $E_{5/2}$, $E_{3/2}$, and $\mu$ 
distinguishes between the two components of each one of the doublets. The quantum numbers $\Omega$ and 
$\mu$ can be determined by considering the operator $\hat T_z$ (called triplex operator in Ref.~\cite{Dudek3}) 
which is the product of a rotation over $\pi/3$ about the symmetry axis followed by the inversion, or parity 
(see Fig.~\ref{Triplex})
\ba
\hat T_z \;=\; \hat P \hat R_z(\pi/3) \;=\; \hat P \, \mbox{e}^{i\pi J_z/3} ~.
\ea
Since the operator $\hat T_z$ is one of the symmetry elements it can be used to define the 
value of $\mu$ as
\ba
\hat T_z \, |\Omega,\mu \rangle \;=\; \mbox{e}^{i\pi \mu/3} \, |\Omega,\mu \rangle ~. 
\label{mu3}
\ea
For even systems of fermions, the identity is given by $(\hat T_z)^6 = 1$, whereas for odd systems, 
due to the double valuedness, one has 
\ba
(\hat T_z)^{12} \;=\; 1 ~,
\ea 
which means that the value of $\mu$ is half-integer and determined up to modulo 6. 
Without loss of generality, one can take $\mu=\pm \frac{1}{2}$, $\pm \frac{3}{2}$, $\pm \frac{5}{2}$. 

The intrinsic state $| \Omega,\mu \rangle$ can be expanded into a basis of states with good angular momentum 
and parity as (see Table~\ref{d3hj}) 
\ba
| \Omega,\mu \rangle \;=\; \sum_{J^P K} B^{\Omega \mu}_{J^P K} \, | J^P K \rangle ~, 
\ea
to obtain
\ba
\hat T_z \, |\Omega,\mu \rangle \;=\; \sum_{J^P K} B^{\Omega \mu}_{J^P K} \, 
P \, \mbox{e}^{i\pi K/3} \, | J^P K \rangle ~,
\ea
which, in combination with Eq.~(\ref{mu3}), implies the following relation between $K$, $P$ and $\mu$ 
\ba
\mu \;=\; \left\{ \begin{array}{ccc} K \, (\mbox{mod }6) &\hspace{0.5cm}& P=+ \\ && \\
(K+3) (\mbox{mod }6) && P=- \end{array} \right.
\label{muK3}
\ea
The values of $\mu$ of each spinor representation $\Omega$ can be determined as follows. 
According to Table~\ref{d3hj}, the angular momentum state $J^P=1/2^+$ has $\Omega=E_{1/2}$, 
and therefore $\mu=K=\pm 1/2$. Similarly, the $J^P=1/2^-$ state has $\Omega=E_{5/2}$, and 
the projections $K=\pm 1/2$ give rise to $\mu=\mp 5/2$. Finally, the $J^P=3/2^+$ state 
decomposes into $E_{1/2}$ with $K=\mu=\pm 1/2$, and $E_{3/2}$ with $K=\mu=\pm 3/2$. 
This shows that the eigenvalues of $\hat T_z$ can be used to identify the triangular symmetry 
of the eigenstates of the CSM. The value of $\mu$ uniquely determines $\Omega$. 
The results are summarized in Table~\ref{lamu3}.  

\begin{table}[h]
\centering
\caption{Eigenvalues of the triplex operator $\hat T_z$, and the classification of basis states 
with triangular symmetry.}
\label{lamu3}
\begin{tabular}{cccc}
\hline
\noalign{\smallskip}
& & \multicolumn{2}{c}{$m=K$} \\ 
$| \Omega,\mu \rangle$ & $\langle \Omega,\mu \, | \, \hat T_z \, | \, \Omega,\mu \rangle$ & $P=+$ & $P=-$ \\
\noalign{\smallskip}
\hline
\noalign{\smallskip}
$| E_{1/2},\pm \frac{1}{2} \rangle$ & $+\frac{1}{2} \sqrt{3} \pm \frac{1}{2}i$ 
& $\pm \frac{1}{2}+6\kappa$ & $\mp \frac{5}{2}+6\kappa$ \\
\noalign{\smallskip}
$| E_{5/2},\pm \frac{5}{2} \rangle$ & $-\frac{1}{2} \sqrt{3} \pm \frac{1}{2}i$ 
& $\pm \frac{5}{2}+6\kappa$ & $\mp \frac{1}{2}+6\kappa$ \\
\noalign{\smallskip}
$| E_{3/2},\pm \frac{3}{2} \rangle$ & $\pm i$ 
& $\pm \frac{3}{2}+6\kappa$ & $\mp \frac{3}{2}+6\kappa$ \\
\noalign{\smallskip}
\hline
\end{tabular}
\end{table}

Moreover, the relation between $\mu$, $K$ and $P$ of Eq.~(\ref{muK3}) allows to split the spherical 
single-particle basis $|nljm=K\rangle$ of the CSM into six sets of basis states according to the value of $\mu$. 
The results are presented in the last two columns of Table~\ref{lamu3} (with $\kappa=0,\pm 1,\ldots$). 
For example, in a calculation in the CSM with a maximum of two oscillator shells the single-particle 
orbits are $1s_{1/2}$, $1p_{1/2}$, $1p_{3/2}$, $2s_{1/2}$, $1d_{3/2}$, $1d_{5/2}$. The basis for each 
one of the intrinsic states $| \Omega,\mu \rangle$ is given by 
\ba
\begin{array}{ccl}
| E_{1/2},\pm \frac{1}{2} \rangle &:\qquad& 1s_{\frac{1}{2},\pm \frac{1}{2}}, \, 2s_{\frac{1}{2},\pm \frac{1}{2}}, 
\, 1d_{\frac{3}{2},\pm \frac{1}{2}}, \, 1d_{\frac{5}{2},\pm \frac{1}{2}} \\ && \\ 
| E_{5/2},\pm \frac{5}{2} \rangle &:\qquad& 1d_{\frac{5}{2},\pm \frac{5}{2}}, \, 1p_{\frac{1}{2},\mp \frac{1}{2}}, 
\, 1p_{\frac{3}{2},\mp \frac{1}{2}} \\ && \\
| E_{3/2},\pm \frac{3}{2} \rangle &:\qquad& 1d_{\frac{3}{2},\pm \frac{3}{2}}, \, 1d_{\frac{5}{2},\pm \frac{3}{2}}, 
\, 1p_{\frac{3}{2},\mp \frac{3}{2}}
\end{array}
\ea

The triangular symmetry requires the wave function to be invariant under the action of 
the (simplex) operator consisting of the product of a rotation over $\pi$ about 
an axis perpendicular to the symmetry axis followed by an inversion. Without loss of 
generality we take a rotation about the $y$-axis 
\ba
\hat S_y \;=\; \hat P_L \, \mbox{e}^{i\pi L_2}
\;=\; \hat P \, \mbox{e}^{i\pi J_2} \, \hat p \, \mbox{e}^{-i\pi j_2} ~.
\label{Sy}
\ea
As a result, the wave function can be written as the product of an intrinsic and a 
collective part
\ba
| \Omega,\mu; J^P KM \rangle \;=\; \frac{1}{\sqrt{2}} 
\left( 1 + \hat P \, \mbox{e}^{i\pi J_2} \, \hat p \, \mbox{e}^{-i\pi j_2} \right) 
| J^P KM \rangle \, | \Omega,\mu \rangle ~.
\label{wftriangle}
\ea
Here we have used that for the triangular configuration the projection $K$ of the angular momentum $J$ 
on the symmetry axis is a good quantum number. 

\subsection{Rotational energies}

The calculations for the CSM are carried out in the intrinsic, or body-fixed, frame. 
The rotational energies can be obtained from
\ba
H_{\rm coll} \;=\; \sum_{i=1}^3 \frac{L_i^2}{2{\cal I}_i} 
\;=\; \sum_{i=1}^3 \frac{(J_i - j_i)^2}{2{\cal I}_i} 
\;=\; \sum_{i=1}^3 \frac{J_i^2}{2{\cal I}_i} + \sum_{i=1}^3 \frac{j_i^2}{2{\cal I}_i} 
- \sum_{i=1}^3 \frac{2 J_i j_i}{2{\cal I}_i} ~,
\ea
which for the present case with ${\cal I}_1={\cal I}_2={\cal I} \neq {\cal I}_3$ reduces to
\ba
H_{\rm coll} \;=\; \frac{1}{2{\cal I}} \left[ \vec{J}^2 + \vec{j}^2 - J_3^2 - j_3^2 
- (J_+ j_- + J_- j_+) \right] + \frac{1}{2{\cal I}_3} (J_3 - j_3)^2 ~.
\ea
The second term proportional to $\vec{j}^2$ only depends on single-particle degrees of freedom and 
can be absorbed into the Hamiltonian of Eq.~(\ref{hcsm}), and will not be considered any further. 
The expectation values of $(J_3 - j_3)^2$ and $j_3^2$ can be calculated as 
\ba
\langle (J_3 - j_3)^2 \rangle &=& \sum_{nljm} |C_{nljm}^{\Omega \mu}|^2 (K-m)^2 \;\approx\; 0 ~,
\nonumber\\
\langle j_3^2 \rangle &=& \sum_{nljm} |C_{nljm}^{\Omega \mu}|^2 m^2 \;\approx\; K^2 ~,
\label{aprox}
\ea
where the coefficients $C$ are the expansion coefficients of the intrinsic states 
in the spherical basis 
\ba
| \Omega,\mu \rangle \;=\; \sum_{nljm} C_{nljm}^{\Omega \mu} \, | nljm \rangle ~.
\ea
The approximate values in Eq.~(\ref{aprox}) are obtained in a calculation in which the value of 
$\beta=1.74$ fm was determined from the first minimum in the elastic form factor of $^{12}$C \cite{C12-2}. 

\begin{figure}
\centering
\resizebox{0.75\columnwidth}{!}{\includegraphics{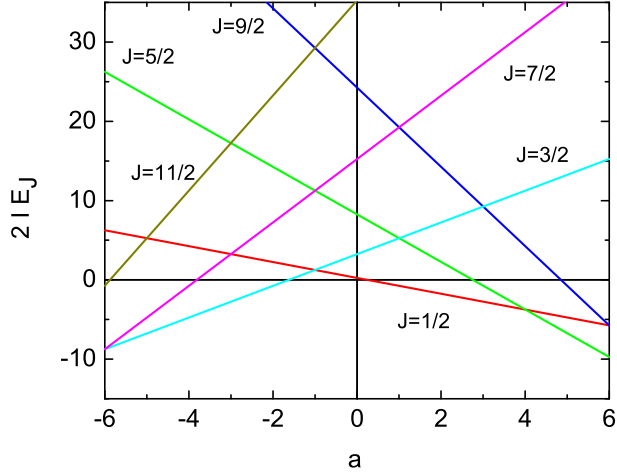}} 
\caption{Dependence of rotational energies on the decoupling parameter $a=a_{\Omega}$.}
\label{Coriolis3}
\end{figure}

The rotational spectrum is given by
\ba
E_{\Omega}(J) \;\approx\; \frac{1}{2{\cal I}} \left[ J(J+1) - 2K^2
+ \delta_{K,1/2} \, a_{\Omega} (-1)^{J+1/2} \left( J+\frac{1}{2} \right) \right] ~,
\label{erot}
\ea
where the last term denotes the Coriolis mixing. 
For the $\Omega=E_{1/2}$ band the decoupling parameter is given by
\ba
a_{E_{1/2}} &=& - \left< E_{1/2},1/2   
\left| \, j_+ \hat{p} \, \mbox{e}^{-i\pi j_2} \right| E_{1/2},1/2 \right>
\nonumber\\
&=& \sum_{nljm} \left| C^{E_{1/2},1/2}_{nljm} \right|^2 
(-1)^{n+j+1/2} \left( j+\frac{1}{2} \right) \delta_{m,1/2} ~,
\ea
and for the $\Omega=E_{5/2}$ band by
\ba
a_{E_{5/2}} &=& \left< E_{5/2},-5/2  
\left| \, j_+ \hat{p} \, \mbox{e}^{-i\pi j_2} \right| E_{5/2},-5/2 \right>
\nonumber\\
&=& - \sum_{nljm} \left| C^{E_{5/2},-5/2}_{nljm} \right|^2 
(-1)^{n+j+1/2} \left( j+\frac{1}{2} \right) \delta_{m,1/2} ~.
\ea
The difference in sign in the expressions for the decoupling parameter is due to the parity 
of the $K=1/2$ bands which is positive for $\Omega=E_{1/2}$ and negative for $\Omega=E_{5/2}$. 
For the $\Omega=E_{3/2}$ band there is no Coriolis mixing, 
\ba
a_{E_{3/2}} \;=\; 0 ~.
\ea
In Fig.~\ref{Coriolis3} we show the dependence of the rotational energies on the decoupling 
parameter in a $K=1/2$ band, as calculated with Eq.~(\ref{erot}).

\section{Tetrahedral symmetry: $T_d$}

For the case of four identical $\alpha$-particles with tetrahedral symmetry 
we choose the $z$-axis along one of the three rotation-inversion axes. 
In this geometry, the coordinates of the three $\alpha$-particles are given by 
$\vec{r}_1=(\beta,\theta,0)$, $\vec{r}_2=(\beta,\theta,\pi)$, 
$\vec{r}_3=(\beta,\pi-\theta,\pi/2)$ and $\vec{r}_3=(\beta,\pi-\theta,3\pi/2)$, 
with $\cos \theta = 1/\sqrt{3}$. Note, that this configuration is different 
from the one used in Ref.~\cite{CSM1}. With this choice, the geometric factor 
in the Hamiltonian becomes
\ba
\sum_{i=1}^{4} Y^{\ast}_{\lambda \nu}(\theta_i,\phi_i) 
\;=\; 2 \delta_{\nu,2\kappa} \, Y_{\lambda \nu}(\theta,0) \left[ 1 + (-1)^{\lambda+\kappa} \right] ~,
\label{tetrahedron}
\ea
{\it i.e.} $\nu$ is even. The structure of the rotational bands built on top of each one of the 
intrinsic states can be obtained by studying the point group symmetry of the problem. 

For the case of tetrahedral symmetry the eigenstates of the CSM can be classified according 
to the irreducible representations (irreps) of the double point group ${\cal T}'_{d}$. 
The double group ${\cal T}'_{d}$ has three spinor representations, denoted by Koster in 
applications to crystal physics as $\Gamma_{6}$, $\Gamma_{7}$, $\Gamma_{8}$ \cite{Koster} 
and by Herzberg in applications to molecular physics as $E_{1/2}$, $E_{5/2}$, $G_{3/2}$ 
\cite{Herzberg3}, and in analogy with \cite{BI2019} in applications to nuclear physics as 
$E_{1/2}^{(+)} \equiv \Gamma_{6} \equiv E_{1/2}$, 
$E_{1/2}^{(-)} \equiv \Gamma_{7} \equiv E_{5/2}$ and $G_{3/2} \equiv \Gamma_{8} \equiv G_{3/2}$. 
The first two are doubly degenerate, whereas the last one has a fourfold degeneracy. 
Again we use the notation introduced by Herzberg.

Table~\ref{tdj} shows the resolution of angular momentum states into irreps of the double point 
group ${\cal T}'_{d}$. As in the previous example, the resolution depends on angular momentum $J$
and parity $P$. The results for higher angular momenta ($J \ge 25/2$) are given by  
\ba
D_J^P \;=\; D_{J-12}^P + (2E_{1/2}+2E_{5/2}+4G_{3/2}) ~.
\ea
The rotational states for each one of the irreps of $T_d$ are given by (by reading Table~\ref{tdj} 
``vertically'')
\ba
\begin{array}{lcl}
\Omega = E_{1/2} &: \qquad& J^P = \frac{1}{2}^+, \frac{5}{2}^-, \frac{7}{2}^{\pm}, 
\frac{9}{2}^+, \frac{11}{2}^{\pm}, \ldots \\ && \\
\Omega = E_{5/2} &: \qquad& J^P = \frac{1}{2}^-, \frac{5}{2}^+, \frac{7}{2}^{\pm}, 
\frac{9}{2}^-, \frac{11}{2}^{\pm}, \ldots \\ && \\
\Omega = G_{3/2} &: \qquad& J^P = \frac{3}{2}^{\pm}, \frac{5}{2}^{\pm}, \frac{7}{2}^{\pm}, 
(\frac{9}{2}^{\pm})^2, (\frac{11}{2}^{\pm})^2, \ldots
\end{array}
\ea

\begin{table}[b]
\centering
\caption{Resolution of rotational states with half-integer angular momentum $J$ and parity $P$
into irreps of $T_d$ \cite{Koster,Herzberg3}.
\label{tdj}}
\begin{tabular}{c|ccc|c|ccccc}
\hline
\noalign{\smallskip}
& $E_{1/2}^{(+)}$ & $E_{1/2}^{(-)}$ & $G_{3/2}$ & 
& $E_{1/2}^{(+)}$ & $E_{1/2}^{(-)}$ & $G_{3/2}$ & \\ 
\noalign{\smallskip}
$D_J^P$ & $\Gamma_6$ & $\Gamma_7$ & $\Gamma_8$ &
$D_J^P$ & $\Gamma_6$ & $\Gamma_7$ & $\Gamma_8$ & \cite{Koster} \\
\noalign{\smallskip}
& $E_{1/2}$ & $E_{5/2}$ & $G_{3/2}$ & 
& $E_{1/2}$ & $E_{5/2}$ & $G_{3/2}$ & \cite{Herzberg3} \\ 
\noalign{\smallskip}
\hline
\noalign{\smallskip}
$1/2^+$ & 1 & 0 & 0 & $1/2^-$ & 0 & 1 & 0 & \\
\noalign{\smallskip}
$3/2^+$ & 0 & 0 & 1 & $3/2^-$ & 0 & 0 & 1 & \\
\noalign{\smallskip}
$5/2^+$ & 0 & 1 & 1 & $5/2^-$ & 1 & 0 & 1 & \\
\noalign{\smallskip}
$7/2^+$ & 1 & 1 & 1 & $7/2^-$ & 1 & 1 & 1 & \\
\noalign{\smallskip}
$9/2^+$ & 1 & 0 & 2 & $9/2^-$ & 0 & 1 & 2 & \\
\noalign{\smallskip}
$11/2^+$ & 1 & 1 & 2 & $11/2^-$ & 1 & 1 & 2 & \\
\noalign{\smallskip}
$13/2^+$ & 1 & 2 & 2 & $13/2^-$ & 2 & 1 & 2 & \\
\noalign{\smallskip}
$15/2^+$ & 1 & 1 & 3 & $15/2^-$ & 1 & 1 & 3 & \\
\noalign{\smallskip}
$17/2^+$ & 2 & 1 & 3 & $17/2^-$ & 1 & 2 & 3 & \\
\noalign{\smallskip}
$19/2^+$ & 2 & 2 & 3 & $19/2^-$ & 2 & 2 & 3 & \\
\noalign{\smallskip}
$21/2^+$ & 1 & 2 & 4 & $21/2^-$ & 2 & 1 & 4 & \\
\noalign{\smallskip}
$23/2^+$ & 2 & 2 & 4 & $23/2^-$ & 2 & 2 & 4 & \\
\noalign{\smallskip}
$25/2^+$ & 3 & 2 & 4 & $25/2^-$ & 2 & 3 & 4 & \\
\noalign{\smallskip}
\hline
\end{tabular}
\end{table}

\begin{figure}
\centering
\resizebox{0.9\columnwidth}{!}{\includegraphics{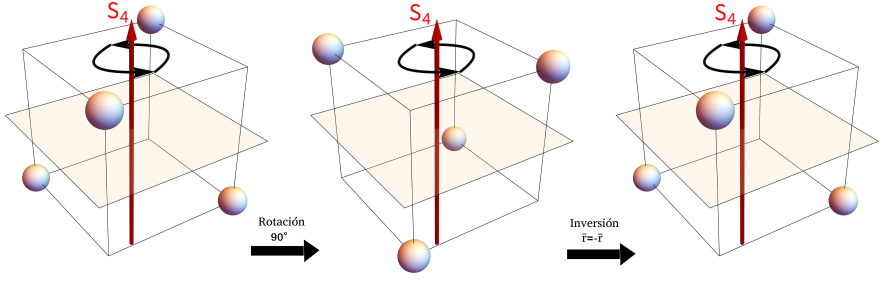}} 
\caption{Doublex symmetry.}
\label{Doublex}
\end{figure}

The CSM eigenstates do not have good angular momentum nor parity, but can be classified 
according to the irreps of the ${\cal T}'_{d}$ tetrahedral symmetry as $|\Omega,\mu \rangle$, 
where $\Omega$ denotes each one of the spinor representations $\Omega=E_{1/2}$, $E_{5/2}$, 
$G_{3/2}$, and $\mu$ distinguishes between the two components of each one of the doublets, 
$E_{1/2}$ and $E_{5/2}$, and the four components of the quartet, $G_{3/2}$.  
The transformation character of the CSM eigenstates can be determined by considering the 
(doublex) operator $\hat D_z$ \cite{Schunck} which is the product of a rotation over 
$\pi/2$ about the symmetry axis followed by the inversion, or parity (see Fig.~\ref{Doublex}) 
\ba
\hat D_z \;=\; \hat P \, \mbox{e}^{i\pi J_z/2} ~.
\ea
Since the operator $\hat D_z$ is one of the symmetry elements it can be used to define the 
value of $\mu$ as
\ba
\hat D_z \, |\Omega,\mu \rangle \;=\; \mbox{e}^{i\pi \mu/2} \, |\Omega,\mu \rangle ~.
\label{mu4}
\ea
In this case, for even systems of fermions the identity is given by $(\hat D_z)^4 = 1$, 
whereas for odd systems, due to the double valuedness, one has 
\ba
(\hat D_z)^{8} \;=\; 1 ~,
\ea 
which means that the value of $\mu$ is half-integer and determined up to modulo 4. 
Without loss of generality, one can take $\mu=\pm \frac{1}{2}$, $\pm \frac{3}{2}$. 

The intrinsic state $| \Omega,\mu \rangle$ can be expanded into a basis of states with 
good angular momentum and parity (see Table~\ref{tdj}) to obtain
\ba
\hat D_z \, |\Omega,\mu \rangle \;=\; \sum_{J^P K} B^{\Omega \mu}_{J^P K} \, 
P \, \mbox{e}^{i\pi K/2} \, | J^P K \rangle ~,
\ea
which, in combination with Eq.~(\ref{mu4}), implies the following relation between $K$, $P$ and $\mu$ 
\ba
\mu \;=\; \left\{ \begin{array}{ccc} K \, (\mbox{mod }4) &\hspace{0.5cm}& P=+ \\ && \\
(K+2) (\mbox{mod }4) && P=- \end{array} \right.
\label{muK4}
\ea
The allowed values of $\mu$ for each spinor representation $\Omega$ can be determined with the help 
of Table~\ref{tdj}. The angular momentum states $J^P=1/2^+$ and $1/2^-$ have $E_{1/2}$ and 
$E_{5/2}$ symmetry, respectively. Their projections $K=\pm 1/2$ correspond to $\mu=\pm 1/2$ 
for positive parity and $\mu=\mp 3/2$ for negative parity. The $J^P=3/2^+$ state has $G_{3/2}$ 
symmetry with $K=\mu=\pm 1/2$, $\pm 3/2$. Therefore, the doublex operator $\hat D_z$ can be used 
to determine the value of $\mu$, in the same way as the triplex operator $\hat T_z$ for the case 
of triangular symmety. However, for the case of tetrahedral symmetry this is not sufficient to 
determine the value of $\Omega$, since the states with $\Omega=E_{1/2}$ and $G_{3/2}$ both have 
components with $\mu=\pm 1/2$, and similarly the states with $\Omega=E_{5/2}$ and $G_{3/2}$ both 
have $\mu=\pm 3/2$. The value of $\Omega$ can be determined by considering the doublex operators 
about the other rotation-inversion axes, 
\ba
\hat D_{x \pm y} \;=\; \hat P \, \mbox{e}^{i \frac{\pi}{2} \left( \frac{J_x \pm J_y}{\sqrt{2}} \right)} 
\;=\; \hat P \, R_z(\pm \pi/4) \, R_y(\pm \pi/2) \, R_z(\mp \pi/4) ~.
\ea
The expectation value of these operators satisfies
\ba
\langle \Omega, \mu | \hat D_{x+y} | \Omega, \mu \rangle \;=\;
\langle \Omega, \mu | \hat D_{x-y} | \Omega, \mu \rangle ~.
\ea
The results are summarized in Table~\ref{lamu4}. 

Eq.~(\ref{muK4}) shows that the tetrahedral symmetry makes it possible to split the spherical basis 
$|nljm\rangle$ into four sets of basis states according to the value of $\mu$ which are not mixed 
by the CSM Hamiltonian of Eq.~(\ref{hcsm}).  
The results are presented in the last two columns of Table~\ref{lamu4} (with $\kappa=0,\pm 1,\ldots$). 
For the case of a calculation with a maximum of two oscillator shells the basis for each 
one of the intrinsic states $| \Omega,\mu \rangle$ is given by 
\ba
\begin{array}{ccl}
| \Omega,\pm \frac{1}{2} \rangle &:\qquad& 1s_{\frac{1}{2},\pm \frac{1}{2}}, \, 2s_{\frac{1}{2},\pm \frac{1}{2}}, 
\, 1d_{\frac{3}{2},\pm \frac{1}{2}}, \, 1d_{\frac{5}{2},\pm \frac{1}{2}}, \, 1p_{\frac{3}{2},\mp \frac{3}{2}} \\ && \\ 
| \Omega,\pm \frac{3}{2} \rangle &:\qquad& 1d_{\frac{3}{2},\pm \frac{3}{2}}, \, 1d_{\frac{5}{2},\pm \frac{3}{2}}, 
\, 1d_{\frac{5}{2},\mp \frac{5}{2}}, \, 1p_{\frac{1}{2},\mp \frac{1}{2}}, \, 1p_{\frac{3}{2},\mp \frac{1}{2}}
\end{array}
\ea
In this case, the value of $\Omega$ of a given eigenstate can be determined by evaluating 
the expectation value of $\hat D_{x \pm y}$ (see Table~\ref{lamu4}).

\begin{table}[h]
\centering
\caption{Expectation values of the doublex operators $\hat D_z$ and $\hat D_{x \pm y}$, 
and classification of basis states with tetrahedral symmetry.}
\label{lamu4}
\begin{tabular}{ccccc}
\hline
\noalign{\smallskip}
&&& \multicolumn{2}{c}{$m=K$} \\
$| \Omega,\mu \rangle$ & $\langle \Omega,\mu \, | \, \hat D_z \, | \, \Omega,\mu \rangle$  
& $\langle \Omega,\mu \, | \, \hat D_{x \pm y} \, | \, \Omega,\mu \rangle$ & $P=+$ & $P=-$ \\
\noalign{\smallskip}
\hline
\noalign{\smallskip}
$| E_{1/2},\pm \frac{1}{2} \rangle$ & $(+1 \pm i)/\sqrt{2}$ & $+1/\sqrt{2}$ 
& $\pm \frac{1}{2}+4\kappa$ & $\mp \frac{3}{2}+4\kappa$ \\
\noalign{\smallskip}
$| E_{5/2},\pm \frac{3}{2} \rangle$ & $(-1 \pm i)/\sqrt{2}$ & $-1/\sqrt{2}$ 
& $\pm \frac{3}{2}+4\kappa$ & $\mp \frac{1}{2}+4\kappa$ \\
\noalign{\smallskip}
$| G_{3/2},\pm \frac{1}{2} \rangle$ & $(+1 \pm i)/\sqrt{2}$ & $-1/2\sqrt{2}$ 
& $\pm \frac{1}{2}+4\kappa$ & $\mp \frac{3}{2}+4\kappa$ \\
\noalign{\smallskip}
$| G_{3/2},\pm \frac{3}{2} \rangle$ & $(-1 \pm i)/\sqrt{2}$ & $+1/2\sqrt{2}$ 
& $\pm \frac{3}{2}+4\kappa$ & $\mp \frac{1}{2}+4\kappa$ \\
\noalign{\smallskip}
\hline
\end{tabular}
\end{table}

The tetrahedral symmetry requires the wave function to be invariant under the action of 
the (simplex) operator of Eq.~(\ref{Sy}), consisting of the product of a rotation over $\pi$ 
about an axis perpendicular to the symmetry axis followed by an inversion \cite{Dudek4}. 
As a result, the wave function for the $\Omega=E_{1/2}$ and $E_{5/2}$ representations 
can be written as the product of an intrinsic and a collective part
\ba
| \Omega,\mu; J^P M \rangle \;=\; \frac{1}{\sqrt{2}} 
\left( 1 + \hat P \, \mbox{e}^{i\pi J_2} \, \hat p \, \mbox{e}^{-i\pi j_2} \right) 
| J^P M \rangle \, | \Omega,\mu \rangle ~.
\label{wftetrahedron}
\ea
Unlike the case for the triangular configuration in Eq.~(\ref{wftriangle}), for the 
tetrahedral configuration the projection of the angular momentum on the symmetry axis 
is not a good quantum number. 

\subsection{Rotational energies}

\begin{figure}
\centering
\resizebox{0.75\columnwidth}{!}{\includegraphics{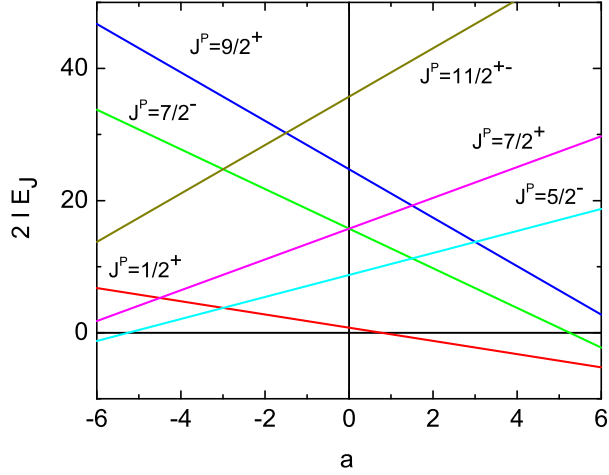}} 
\caption{Dependence of rotational energies on the decoupling parameter $a=a_{E_{1/2}}$.}
\label{Coriolis4}
\end{figure}

The rotational energies can be obtained from
\ba
H_{\rm coll} \;=\; \sum_{i=1}^3 \frac{L_i^2}{2{\cal I}_i} 
\;=\; \sum_{i=1}^3 \frac{(J_i - j_i)^2}{2{\cal I}_i} ~,
\ea
which for the present case with ${\cal I}_1={\cal I}_2={\cal I}_3={\cal I}$ reduces to
\ba
H_{\rm coll} \;=\; \frac{1}{2{\cal I}} \left[ \vec{J}^2 + \vec{j}^2 - 2\vec{J} \cdot \vec{j} \right] ~.
\ea
The recoil term proportional to $\vec{j}^2$ only depends on single-particle degrees of freedom and 
can be absorbed into the CSM Hamiltonian, and will not be considered any further. 
The Coriolis mixing for rotational bands with tetrahedral symmetry was studied in detail in the 
context of mean-field calculations in Ref.~\cite{Dudek4}. 
The rotational spectrum is given by 
\ba
E_{\Omega}(J) \;=\; \frac{1}{2{\cal I}} \left[ J(J+1) + a_{\Omega} g_{\Omega}(J) \right] ~,
\ea
where the second term denotes the Coriolis mixing. 
For the $\Omega=E_{1/2}$ band the decoupling parameter is given by \cite{Dudek4}
\ba
a_{E_{1/2}} &=& \left< E_{1/2},1/2 \left| \, j_3 
- j_+ \hat{p} \, \mbox{e}^{-i\pi j_2} \, \right| E_{1/2},1/2 \right>
\nonumber\\
&=& \sum_{nljm} \left| C^{E_{1/2},1/2}_{nljm} \right|^2 m 
+ \sum_{nlj} \left| C^{E_{1/2},1/2}_{nlj,1/2} \right|^2 (-1)^{n+j+1/2} \left( j+\frac{1}{2} \right) ~,
\ea
with 
\ba
g_{E_{1/2}} \;=\; \left\{ \begin{array}{lll}
-\frac{2}{3}(J+1) && \mbox{ for } J^P= \frac{1}{2}^+, \frac{7}{2}^-, \frac{9}{2}^+, \frac{13}{2}^{\pm}, \ldots \\
&& \\
+\frac{2}{3} J && \mbox{ for } J^P=\frac{5}{2}^-, \frac{7}{2}^+, \frac{11}{2}^{\pm}, \ldots
\end{array} \right.
\ea
In Fig.~\ref{Coriolis4} we show the dependence of rotational energies on the decoupling parameter $a_{\Omega}$ 
for $\Omega=E_{1/2}$.

For the $\Omega=E_{5/2}$ band which is the parity conjugate of the $E_{1/2}$ band, 
the decoupling parameter is given by \cite{Dudek4}
\ba
a_{E_{5/2}} &=& \left< E_{5/2},-3/2 \left| \, j_3 
+ j_+ \hat{p} \, \mbox{e}^{-i\pi j_2} \, \right| E_{5/2},-3/2 \right>
\nonumber\\
&=& \sum_{nljm} \left| C^{E_{5/2},-3/2}_{nljm} \right|^2 m 
- \sum_{nlj} \left| C^{E_{5/2},-3/2}_{nlj,1/2} \right|^2 (-1)^{n+j+1/2} \left( j+\frac{1}{2} \right) ~,
\ea
with
\ba
g_{E_{5/2}} \;=\; \left\{ \begin{array}{lll} -\frac{2}{3}(J+1) && \mbox{ for } 
J^P= \frac{1}{2}^-, \frac{7}{2}^+, \frac{9}{2}^-, \frac{13}{2}^{\pm}, \ldots \\
&& \\
+\frac{2}{3} J && \mbox{ for } J^P=\frac{5}{2}^+, \frac{7}{2}^-, \frac{11}{2}^{\pm}, \ldots
\end{array} \right.
\ea
The dependence of the rotational energies on the decoupling parameter $a_{E_{5/2}}$ is the same 
as for the parity-conjugate band with $\Omega=E_{1/2}$, as shown in Fig.~\ref{Coriolis4}, but for 
rotational states with opposite parities.

For the $\Omega=G_{3/2}$ band the situation is more complicated and, up to now, no analytic expression 
is available. An approximate solution was given in Ref.~\cite{Dudek4}. 

\section{Summary and Conclusions}

We discussed the classification of the eigenstates of the CSM Hamiltonian for the cases of a 
triangular and tetrahedral configuration of $\alpha$-particles. The eigenstates are characterized by 
the irreducible representations $\Omega$ of the double point groups ${\cal D}'_{3h}$ 
and ${\cal T}'_d$, respectively, and a label $\mu$ to distinguish between the different components 
of $\Omega$. For ${\cal D}'_{3h}$ there are three doubly degenerate spinor representations, 
$\Omega=E_{1/2}$, $E_{5/2}$ and $E_{3/2}$, whereas for ${\cal T}'_d$ there are two doubly degerate 
spinor representations $\Omega=E_{1/2}$, $E_{5/2}$ and one with fourfold degeneracy $\Omega=G_{3/2}$. 

In general, in the CSM the eigenstates are obtained numerically by diagonalization of the Hamiltonian 
with the appropriate symmetry. We showed that the discrete symmetry of the eigenstates can be determined 
by a suitable choice of the coordinate system in combination with the expectation value of the triplex 
and doublex operators defined with respect to the rotation-inversion axes. 

However, we showed that for both the triangular ${\cal D}'_{3h}$ and the tetrahedral ${\cal T}'_d$ 
symmetries it is possible to construct a symmetry-adapted basis according to the value of $\mu$ by 
an appropriate choice of single-particle states. For the case of triangular symmetry this is sufficient, 
since the value of $\mu$ uniquely determines $\Omega$. For the tetrahedral symmetry, this is not the 
case, but the value of $\Omega$ can be determined by the evaluating the expectation value of the 
doublex operators. 

Finally, as an application, we derived closed expressions for the Coriolis mixing matrix elements. 

\section*{Acknowledgements}

This work was supported in part by grants IN109017 and IN101320 from DGAPA-UNAM, Mexico (RB), 
and 784896 from CONACyT, Mexico (AHSV).

\end{document}